\documentclass[pre,aps,twocolumn]{revtex4}
\usepackage[latin1]{inputenc}
\usepackage[english]{babel}
\usepackage{graphicx}
\usepackage{color}
\usepackage{amsmath}
\begin{document}

\title{\bf\Large A Langevin canonical approach to the dynamics of
chiral two level systems. Thermal averages and heat capacity}

\author{H. C. Pe\~nate-Rodr\'{\i}guez $^{a,b}$
A. Dorta-Urra$^{c}$, P. Bargue\~no $^{*,d}$, G.
Rojas-Lorenzo$^{a,b}$ and S. Miret-Art\'es$^{**,b}$}
\affiliation{$^{a}$ Instituto Superior de Tecnolog\'{\i}a y
Ciencias Aplicadas, Ave. Salvador Allende y Luaces, Quinta de Los
Molinos, Plaza, La Habana 10600, Cuba
\\
$^{b}$ Instituto de F\'{\i}sica Fundamental (CSIC), Serrano 123
{\it E-28006, Madrid, Spain, ($^{**}$s.miret@iff.csic.es) }
\\
$^{c}$ Unidad asociada UAM-CSIC, Instituto de F\'{\i}sica
Fundamental (CSIC), Serrano 123 {\it E-28006, Madrid, Spain}
\\
$^{d}$ Departamento de F\'{\i}sica de Materiales, Universidad
Complutense de Madrid, {\it E-28040}, Madrid, Spain
($^*$p.bargueno@fis.ucm.es) }

\date{\today}

\begin{abstract}
A Langevin canonical framework for a chiral two--level system
coupled to a bath of harmonic oscillators is developed within a
coupling scheme different to the well known spin-boson model.
Thermal equilibrium values are reached at asymptotic times by
solving the corresponding set of non--linear coupled equations in
a Markovian regime. In particular, phase difference thermal values
(or, equivalently, the so--called coherence factor) and heat
capacity through energy fluctuations are obtained and discussed in
terms of tunneling rates and asymmetries.
\end{abstract}
\maketitle

\section{Introduction}
Two level systems (TLS) can be considered as a paradigmatic
approach found in many different areas of physics and chemistry
going, for example, from chiral molecules
\cite{Harris1978,Quack1986}, electron transfer reactions
\cite{Weiss1999,Transferbook}), quantum optics \cite{Scullybook}
to quantum computation \cite{Ladd2010}, to name only a few of
them. Isolated (two level) systems
are very rare in nature and they are very often coupled to a more
extended system or thermal bath consisting of many degrees of
freedom usually represented by an infinite set of harmonic
oscillators. In general, the TLS is linearly coupled to the
coordinates of a bath of noninteracting oscillators, whose
properties are encoded in their spectral density
\cite{Leggett1987}.
The standard model for such a description is the well--known
spin--boson model \cite{Weiss1999}.
This thermal bath can also be seen as an
independent set of identical systems surrounding the tagged one.
Within the density matrix formalism, path--integral methods and
Bloch--Redfield equations have been proposed and implemented to
determine the time evolution of the dissipative TLS
\cite{Redfield1957,Senitzki1963,Harris1981,Harris1983,Hartmann2000,
Vincenzo2005,Isabel2011}. Variational calculations have also been
carried out for both the symmetric (or unbiased) \cite{Silbey1984}
and asymmetric (or biased) cases \cite{Harris1985}. Alternatively,
according to Feynman in his dynamical theory of the Josephson
effect \cite{Feynman}, it was shown that classical and quantum
mechanics may be embedded in the same Hamiltonian formulation by
using canonical complex coordinates
\cite{Strocchi1966,Heslot1985}. This procedure was ulteriorly used
by Meyer and Miller \cite{Meyer1979} starting from an earlier work
by McCurdy and Miller \cite{McCurdy1978} for  electronically
non--adiabatic processes. In this work, each electronic state was
represented by a pair of classical action-angle variables.
In fact, the Meyer--Miller--Stock--Thoss Hamiltonian
\cite{Meyer1979,Stock1997} for a TLS can be rigourously defined
after introducing appropriate action angle--coordinates on $S^{2}$
\cite{Chruszinskybook}, which can be taken to be the true quantum
phase space where the dynamics takes place
\cite{Pedro2013a,Pedro2013b}. Moreover, it can be shown
\cite{Chruszinskybook} that the dynamics derived from this
classical formulation is completely equivalent to the quantum one
not only for TLS but also for n--level systems.

Very recently, this formalism has also been implemented within the
Langevin canonical framework to study the dissipative and
stochastic dynamics of chiral molecules and TLSs in general
\cite{Bargueno2011,Penate2012,Dorta2012,chirality,Pedro2013a,Pedro2013b}.
In this case, the asymmetry of the assumed double well potential
model is due to the parity violating energy difference. In
particular, the time evolution of the non--isolated chiral TLS has
provided thermal average population difference and coherences for
incoherent and coherent tunneling. This dynamics is discussed in
terms of a critical temperature defined by the maximum of the heat
capacity \cite{Bargueno2009,chirality}. It has been proved that
this approach is able to reproduce path--integral results beyond
the so--called non--interacting blip approximation (NIBA) for a
large range of temperatures and by assuming Ohmic friction
\cite{chirality}. Below this critical temperature is where quantum
effects are much more important than thermal effects and the
coherent regime is well established.

This work can be considered as a next step forward into a more
complete dynamical analysis of this open quantum system by
focusing on different thermodynamical quantities such as average
energies, phase differences and heat capacities issued from a
stochastic dynamics. We are not going to consider here a very
important topic about anomalies of the heat capacity and refer the
reader to some pertinent works
\cite{Weiss1988,Hanggi2008,Ingold2009,Ingold2012}. Thus, Section
II is devoted to briefly introduce the canonical formalism for the
isolated and non--isolated TLS. Finite coupling and temperature
effects are included by means of noise--induced dynamics via a
Caldeira--Leggett--like Hamiltonian. The coupling model is
different to the standard spin-boson model since it is the phase
difference which is responsible for the coupling with the bath
oscillators (in a similar way to the Josephson dynamics).
Numerical simulations of this stochastic approach are presented
and discussed in Section III, where thermal average energy values
as well as phase differences or the so--called coherence factor
\cite{Stringari2001} are evaluated by assuming an Ohmic regime in
a broad range of temperatures. Furthermore, the energy
fluctuations expressed in terms of the heat capacity are also
showed corroborating the maximum found at the critical
temperature. This thermodynamical information is obtained from the
stochastic dynamics at asymptotic times where the thermal
equilibrium is reached. Derived magnitudes of the energy
fluctuations are also the geometric phase and the interference
pattern due to the presence of the two enantiomers. In an
Appendix, we have also shown that if the coupling with the bath is
through the other canonical variable, similar to the spin-boson
model, the same asymptotic behavior is clearly obtained but
following a different time evolution.

\section{A Langevin canonical formalism for the stochastic dynamics of
chiral two level systems}

The isolated TLS is usually modelled by a two-well (asymmetric)
potential within the Born-Oppenheimer approximation and described
by the phenomenological Hamiltonian
\begin{equation}
\hat H=\delta \hat\sigma_{x}+\epsilon \hat\sigma_{z}
\label{Hhat}
\end{equation}
where $\sigma_{x,z}$ stand for the Pauli matrices, $\delta$
accounts for the tunneling rate and $\epsilon$ for the asymmetry
of the potential wells due to electroweak parity violation (for a
chiral system) or any other biasing term (for example, a polarized electric or
magnetic field). In terms of the left and right states (or chiral states),
$|L\rangle$ and $|R\rangle$, respectively, these two parameters
are given by the matrix elements $\langle L | \hat H |R \rangle =-
\delta$ (with $\delta > 0$) and $2\epsilon=\langle L|\hat
H|L\rangle -\langle R|\hat H|R\rangle $ ($\epsilon$ can be
positive or negative).

The time evolution of this system is given by solving  the
time-dependent Sch\"odinger equation ($\hbar=1$)
\begin{equation}
i\,\partial_{t}|\Psi(t)\rangle=\hat H|\Psi(t)\rangle
\label{SE}
\end{equation}
where the total wave function can be expressed in terms of the
chiral states as
\begin{equation}
|\Psi(t)\rangle=a_{L}(t)|L\rangle+a_{R}(t)|R\rangle  .
\end{equation}
Now, if the complex coefficients are written in polar form as
$a_k(t)=|a_k(t)|e^{i\Phi_k(t)}$, where $k$ stands for $R$ or $L$,
and the population and phase differences between chiral states are
defined as $z(t)\equiv|a_{R}(t)|^{2}-|a_{L}(t)|^{2}$ and
$\Phi(t)\equiv\Phi_{R}(t)-\Phi_{L}(t)$, respectively, the average
energy in the normalized $|\Psi(t)\rangle$ state is given by
\begin{equation}
 H_{0} \equiv \langle \Psi|\hat H|\Psi\rangle=
-2\delta\sqrt{1-z^{2}}\cos\Phi+2\epsilon z
\end{equation}
$H_{0}$ being a Hamiltonian function \cite{Quantumbook}. Since $z$
and $\Phi$ are a pair of canonically conjugate variables, the
Hamilton or Heisenberg equations of motion are derived from $\dot
{z} = -\partial H_{0} / \partial \Phi$ and $\dot {\Phi} = \partial
H_{0} / \partial z$ to give
\begin{eqnarray}
\label{TLQSeq}
\dot z &=&-2\delta\sqrt{1-z^2}\sin \Phi \nonumber
\\
\dot \Phi &=&2\delta\frac{z}{\sqrt{1-z^2}}\cos \Phi+
2\epsilon.
\end{eqnarray}
Thus, these two non-linear coupled equations  are equivalent to
Eq. (\ref{SE}). 
For practical purposes, the adimensional time $t\rightarrow
2\delta\, t$ is frequently used in this context. Thus, the
corresponding time scaling implies that the dimensionless
Hamiltonian function $H_{0}$ is now expressed as
\begin{equation}
H_{0}=-\sqrt{1-z^{2}}\cos\Phi+\frac{\epsilon}{\delta} z.
\label{H}
\end{equation}
Note that the first term of the Hamiltonian function (\ref{H})
accounts for the tunneling process due to the presence of the
oscillatory function and, the second one, for the underlying
asymmetry, stressing the fact that two competing processes are
present in this simple dynamics. In addition, the ratio between
$\epsilon$ and $\delta$ is critical in this dynamics since it
provides a way to control the effects of
delocalization/localization. In other words, when the tunneling
rate is much greater than the bias, the first term of the
Hamiltonian (\ref{H}) predominates and an important delocalization
is expected to be present. On the contrary, important asymmetries
localize the dynamics in one of the two potential wells.

When dealing with environment interactions consisting of a high
number of degrees of freedom, several theoretical treatments are
widely used. They are typically classified according to one of the
three pictures of quantum mechanics
\cite{Weiss1999,Petruccione2006}: the density matrix formalism and
the stochastic Schr\"odinger equation (interaction and
Schr\"odinger pictures), and the generalized Langevin equation
(Heisenberg picture). Within our theoretical scheme, it is clear
that the time evolution of the non-isolated, chiral TLS system has
to be carried out in the last picture \cite{Sanzbook1,Sanzbook2}.
In this canonical formalism, a Caldeira--Leggett--like
Hamiltonian, \cite{Leggett1987} where a bilinear coupling between
the TLS and the environment is assumed, has been recently
developed to study the corresponding dissipative and stochastic
dynamics \cite{Bargueno2011,Penate2012,Dorta2012,chirality}. As
stated before, noting that $\Phi$ and $z$ play the role of a
generalized coordinate and momentum, respectively, one can
straightforwardly derive
%
%
the following system of coupled Langevin-type dynamical equations
\cite{Bargueno2011,Penate2012,Dorta2012,chirality}
\begin{eqnarray}
\label{stochohm} \dot z &=&-\sqrt{1-z^2}\sin \Phi \nonumber
\\
&-&\int^{t}_{0}\gamma(t-t') \dot \Phi(t') \, dt' + \xi(t)
\nonumber \\
\dot \Phi &=&\frac{z}{\sqrt{1-z^2}}\cos \Phi+
\frac{\epsilon}{\delta} ,
\end{eqnarray}
where $\gamma (t)$ is the time-dependent friction (or damping
kernel)
%
%
and $\xi(t)$ the fluctuation force or noise.
%
%
Note that this approach does not correspond to the standard
spin-boson model \cite{Weiss1999}.
It follows quite closely that employed in the field of condensed
matter, in particular, the dynamics of a Josephson junction
\cite{Weiss1999}. This phase difference is coupled to the degrees
of freedom of the degree of the bath which also acts as a source
of phase fluctuations. The thermal average of the corresponding
cosine function is called the coherence factor
\cite{Stringari2001} which provides the degree of coherence of the
process. In the Appendix, and for completeness, a coupling
following the standard spin-boson model is also discussed.

When a Markovian regime is assumed, the usual properties of the
fluctuation force (Gaussian white noise) are given by the
following canonical thermal averages: $\langle
\xi(t)\rangle_{\beta}=0$ (zero average) and $\langle
\xi(0)\xi(t)\rangle_{\beta}=m k_{B}T\gamma \delta (t)$
(delta-correlated) where $\beta = (k_B T)^{-1}$, $k_B$ being
Boltzmann's constant. The friction is then described by
$\gamma(t)=2\gamma\delta(t)$, where $\gamma$ is a constant and
$\delta(t)$ is Dirac's $\delta$--function (not to be confused with
the $\delta$-parameter describing the tunneling rate). Thus, in
this regime, Eqs. (\ref{stochohm}) read now as follows
\begin{eqnarray}
\label{stochohm1} \dot z &=&-\sqrt{1-z^2}\sin \Phi - \gamma \dot
\Phi(t)  + \xi(t) \nonumber \\
\dot \Phi &=&\frac{z}{\sqrt{1-z^2}}\cos \Phi+
\frac{\epsilon}{\delta} .
\end{eqnarray}
The corresponding solutions provide stochastic  trajectories for
the population and phase differences encoding all the information
on the dynamics of the non-isolated, chiral TLS. Interestingly
enough it is the fact that from Eqs. (\ref{stochohm1}), an
effective Hamiltonian function which explicitly depends on the
friction and noise can be extracted from the Hamilton equations
given by Eq. (\ref{stochohm1}),
\begin{equation}
H_{\gamma,\xi} (z,\Phi) =  -\sqrt{1-z^{2}}\cos \Phi +
\frac{\epsilon}{\delta}z + \gamma \Phi \dot \Phi - \xi \Phi
\label{Hgamma}
\end{equation}
which represents the non--conserved energy of the chiral system
under the presence of the thermal bath and its mutual coupling as
a function of time. This effective energy turns out to be critical
for the evaluation of any thermodynamical function such as, for
example, the heat capacity. In a certain sense, this information
is the alternative way to the more standard one of using the
density matrix an/or partition function for the reduced system. It
gives us a simple method to evaluate time dependent energy
fluctuations.

On the other hand, the connection to the density matrix formalism
is readily obtained from the corresponding matrix elements
expressed as $\rho_{R,R}=|a_{R}|^{2}$, $\rho_{L,L}=|a_{L}|^{2}$,
$\rho_{L,R}=a_{L}a_{R}^{*}$ and $\rho_{R,L}=a_{R}a_{L}^{*}$. Thus,
the time average values of the Pauli operators are given by
\begin{eqnarray}
\label{correspondence} \langle \hat \sigma_{z}\rangle_t&=&
\rho_{R,R}-\rho_{L,L} = z  \nonumber \\
\langle \hat \sigma_{x}\rangle_t&=& \rho_{R,L}+\rho_{L,R} =
-\sqrt{1-z^{2}}\cos\Phi \nonumber \\
\langle \hat \sigma_{y}\rangle_t&=& i \rho_{R,L} - i \rho_{L,R}
=\sqrt{1-z^{2}}\sin\Phi,
\end{eqnarray}
which is consistent with $\langle\hat H\rangle=\delta\langle \hat
\sigma_{x}\rangle+ \epsilon \langle \hat \sigma_{z}\rangle$ and
\begin{equation} \label{nodamping}
\langle \hat \sigma_{x}\rangle_t^2 + \langle \hat
\sigma_{y}\rangle_t^2 + \langle \hat \sigma_{z}\rangle_t^2 \leq 1.
\end{equation}
where the equal sign holds for the isolated system dynamics.


Several comments are in order when solving numerically Eqs.
(\ref{stochohm1}). First, units along this work are dimensionless.
By doing this, we are considering a general dynamics where any
chiral molecule can be represented. For example, if for a given
chiral molecule $\delta = 10^{-4}$ meV, we set this value to be
$1$ after passing the tunneling rate to inverse of atomic units of
time, $3.675 \, 10^{-5}$. In all the calculation, we have further
assumed that $\delta \sim \epsilon$ in order to properly analyze
the competition between tunneling and asymmetry or between
delocalization and localization, as stated before. With the time
step used, $\gamma = 0.1$ or $0.01$ (dimensionless rate) is a good
parameter for the Ohmic friction. When working on thermodynamic
functions, reduced units have been employed, that is, energies and
temperatures are divided by $\Delta$ (where
$\Delta=\sqrt{\delta^{2}+\epsilon^{2}}$). Second, at high
temperatures, $\beta^{-1} \gg \gamma$, thermal effects are going
to be predominant over quantum effects which become relevant, in
general, at times of the order of or less than the so-called
thermal time, $\beta$ (in atomic units). However, at very low
temperatures, $\beta^{-1} \ll \gamma$, the noise is usually
colored and its auto-correlation function is complex, our approach
being no longer valid. In fact, at cold or ultracold regimes, a
chiral two level bosonic system could display condensation as well
as a discontinuity in the heat capacity (reduced temperatures $k_B
T / \Delta \leq 1$) \cite{Bargueno-pccp}. Here, a canonical
(Maxwell--Boltzman) distribution is assumed and only classical
noise is considered since the ultracold regime is not going to be
analyzed. Third, the role of initial conditions has been
extensively discussed in the literature (see, for example,
\cite{Weiss1999,Petruccione2006}); for practical purposes, the
chiral system will be prepared in one of its two states, left or
right ($z(0)= 0.999$ or $-0.999$ in order to avoid initial
singularities, and very far from the equilibrium condition), and
the initial phase difference $\Phi (0)$ will be uniformly
distributed around the interval $[- \pi, \pi]$. Fourth, the
stochastic trajectories issued from solving Eq. (\ref{stochohm1})
are dependent on the four dimensional parameter space
$(\epsilon,\delta,\gamma,T)$. And fifth, when running trajectories
there are some of them visiting "un-physical" regions, that is,
$|z|>1$. This drawback is mainly associated with the intensity of
the noise since, for large values of it (which depends on both the
temperature and the friction coefficient), the stochastic
$z$--trajectories can become unbounded. To overcome this problem,
we have implemented a {\it reflecting} condition such that when
the trajectory reaches $|z| > 1$, we change its value to $2-|z|$.

The general strategy consist of solving the pair of non-linear
coupled equations (\ref{stochohm1}) for the canonical variables
under the action of a Gaussian white noise, which is implemented
by using an Ermak--like approach \cite{Ermak1980,Allenbook}. Note
that in the Langevin--like coupled equations to be solved, the
noise term only appears in the equation of motion of the
$z$--variable. The time step used is $10^{-2}$ (dimensionless
units) for all the cases analyzed. As noted in
\cite{Bargueno2011}, unstable trajectories can also be found for
certain values of $\epsilon$, $\delta$ and $\gamma$ in the simple
case of dissipative but non--noisy dynamics. As this problem
persists in case of dealing with stochastic trajectories, not
every triplet $(\epsilon,\delta,\gamma)$ gives place to a stable
trajectory. In these cases, the time evolution of each individual
trajectory is not possible and a previous stability analysis is
mandatory. However, in the stable case, a satisfactory description
of population differences and coherences have been achieved by
running up to $ 10^{4}$ trajectories as already reported elsewhere
\cite{chirality}.

\section{Thermodynamics from stochastic dynamics}

In general, there are several routes to reach thermodynamical
properties such as partition functions, thermal averages, heat
capacity, entropy, etc. Very likely, the most popular one is that
based on the thermodynamic method coming from the path--integral
formalism. An extensive account of this formalism can be found in
Weiss's book \cite{Weiss1999}. However, numerical instability
problems from the analytic continuation of certain functions lead
to some drawbacks. As an alternative way to avoid such problems,
it can also be proposed the opposite procedure, that is, the
computation of thermodynamics from dynamics which may have some
advantages. Thus, partition functions and canonical thermal
averages are then calculated when carrying out dynamical
calculations for different bias or asymmetry. Analogously, one can
also obtain the main equilibrium thermodynamics properties of the
non-isolated TLS from the stochastic dynamics at asymptotic times
(if the dynamics is ergodic) for a given bias and different
temperatures. Furthermore, it is worth stressing that the
thermodynamic functions are independent on the friction
coefficient in the weak coupling limit. Therefore, our
thermodynamical average values coming of solving the corresponding
stochastic dynamics are independent on the friction coefficient as
time goes to infinity, that is, when the thermal equilibrium with
the bath is reached. In the strong coupling limit, this fact no
longer holds \cite{Ingold2009}.

A previous analysis of the thermodynamics of non-interacting
chiral molecules assuming a canonical distribution has been
carried out elsewhere \cite{Bargueno2009}. In particular, thermal
averages of pseudoscalar operators were extensively analyzed. The
canonical thermal average of an observable $X$ is defined as
$\langle X \rangle_{\beta}= \mathrm{Tr}(\rho_{\beta}X)$ where
$\rho_{\beta}=Z^{-1}e^{-\beta H}$, $H$ being given by
(\ref{Hhat}). The quantum partition function $Z$ is given by
$Z=2\cosh(\beta \Delta)$ from the eigenvalues of $H$. Moreover,
the corresponding averages for the population difference and
coherences are then calculated to give
\begin{eqnarray}\label{thermo}
\langle z \rangle_{\beta} \equiv \langle \hat
\sigma_{z}\rangle_{\beta}&=&\frac{\epsilon}{\Delta}\tanh(\beta
\Delta) \nonumber
\\
\langle \hat \sigma_{x}\rangle_{\beta}&=&\frac{\delta}{\Delta}
\tanh(\beta \Delta)  .
\end{eqnarray}
These populations and coherences have also been evaluated from the
stochastic dynamics leading to numerical values in agreement with
Eqs. (\ref{thermo}) \cite{chirality}. Furthermore, depending on
the temperature, the incoherent and coherent tunneling regimes
were fitted to path--integral analytical expressions beyond the
NIBA in order to extract information of the frequencies and
damping factors of the non--isolated system in its time evolution
to thermodynamic equilibrium. The critical temperature issued from
the maximum of the heat capacity \cite{Bargueno2009} can be
considered the threshold temperature where quantum effects become
dominant; or in other words, where the coherent regime is well
established.

\begin{figure}[h!]
\includegraphics[width=0.5\textwidth,height=0.5\textwidth,angle=-90]
{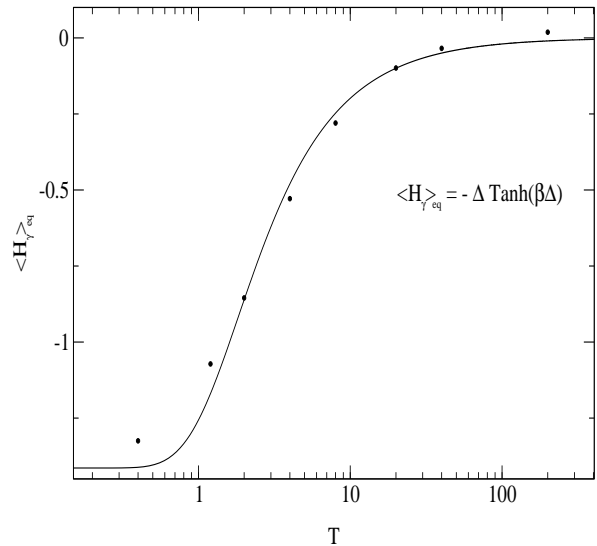} \caption{\label{fig1}  Thermal average of the energy of
the chiral TLS as a function of the temperature (dimensionless
units). Black points are issued from solving the corresponding
non--linear equations of motion, Eqs. (\ref{stochohm1}), and solid
line from the thermodynamics function given by Eq. (\ref{Eave}).
Here, $\delta = \epsilon = 1$ and $\gamma$ is 0.1 for $T
> 1$ and 0.01 for $T < 1$.}
\end{figure}

On the other hand, the thermal average of the energy has been
showed to be \cite{Bargueno2009}
\begin{equation}
\langle E \rangle_{\beta} = E_0 - \Delta \tanh (\beta \Delta)  .
\label{Eave}
\end{equation}
Usually, the origin of the energy is taken to be $E_0 = (\langle
L|\hat H|L\rangle  + \langle R|\hat H|R\rangle) / 2$. Note that
the signature of the thermal average is the global factor $\tanh
(\beta \Delta)$, reminiscence of the eigenvalues of the $H$
Hamiltonian (\ref{Hhat}). This thermodynamical quantity can also
be easily extracted from the time evolution of the chiral system
from the effective Hamiltonian function defined by Eq.
(\ref{Hgamma}) at asymptotic times, that is,
\begin{equation}
\langle E \rangle_{\beta} = \langle  H_{\gamma,\xi} (z,\Phi)
\rangle_{\beta}  \label{Eave-t}
\end{equation}
In Fig. \ref{fig1}, the thermal average of the energy (solid line)
given by Eq. (\ref{Eave}) is plotted versus a wide interval of
temperatures going from 0.1 to 200 in units of $\Delta$ (reduced
temperatures); black points are the asymptotic values of the
non-conserved energy of the chiral system when solving the
equations of motion, Eqs. (\ref{stochohm1}), for values of $\delta
= \epsilon = 1$. The $\gamma$ parameter is fixed at 0.1 for
temperatures $T > 1$ and 0.01 for $T < 1$ in order to facilitate
the calculations. Remember that the equilibrium state is
independent on $\gamma$ for weak coupling with the bath.

\begin{figure}[!h]
\includegraphics[width=0.5\textwidth,height=0.5\textwidth,angle=-90]
{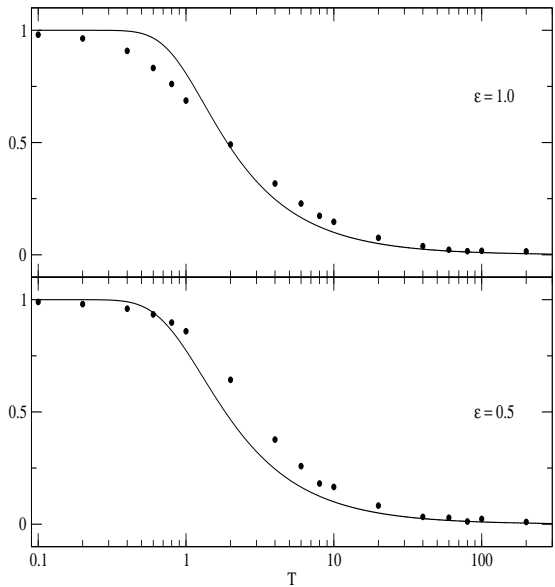} \caption{\label{fig2} Canonical thermal
average of the so--called coherence factor $\langle cos \Phi
\rangle_{\beta}$ as a function of the temperature. Here, $\delta =
1$, $\epsilon = 0.5,1$  and $\gamma$ is 0.1 for $T
> 1$ and 0.01 for $T < 1$.}
\end{figure}

At this point, one could argue that thermal averages could also be
extracted analytically from its definition in standard statistical
mechanics, that is,
\begin{equation}
\langle F(z, \Phi) \rangle_{\beta} = \frac{1}{Z_c} \int_{-1}^{1}
dz \int_{0}^{2 \pi} d \Phi  F(z, \Phi) e^{- \beta H_0 (z,\Phi)}
\label{F}
\end{equation}
where $F(z, \Phi)$ is a general function of the canonical
variables $z$ and $\Phi$ and $H_0$ is the Hamiltonian given by Eq.
(\ref{H}). A straightforward analytical integration of the
corresponding partition function $Z_c$ gives
\begin{equation}
Z_c = \frac{4 \pi}{\beta \Delta} \, \mathrm{sinh} \beta \Delta
\end{equation}
which is quite different from the quantum partition function
mentioned above. The corresponding thermal averages of $z$, $E$,
$\sigma_x$, etc. can also be easily evaluated analytically. The
corresponding results are clearly different from the ones
previously discussed. In principle, one should expect agreement
when the dynamical conditions are approaching those of a classical
system (for example, by increasing the temperature).

Special attention deserves the thermal average of $\Phi$ and $\cos
\Phi$ \cite{Sringari2001}, the last average being also called
coherence factor, $\langle \cos \Phi \rangle_{\beta}$ (which
provides the degree of coherence of the chiral system). In
particular, the quantum thermal average of the coherence factor is
given by
\begin{equation}
\langle \cos \Phi \rangle_{\beta} = \frac{\sum_n \langle n | \cos
\Phi | n \rangle  e^{- \beta E_n}}{\sum_n e^{- \beta E_n}}
\label{cos}
\end{equation}
where  the sum over $n$ runs only two values, the two eigenstates.
The quantum averages of $\cos \Phi$ over these two eigenstates
could be carried out following Ref. \cite{Carruthers1968}. Here,
however, we are going to follow a simpler procedure. Due to the
fact the equilibrium values of $\langle z \rangle_{\beta}$ and
$\langle \sigma_x \rangle_{\beta}$, Eq. (\ref{thermo}), are well
reproduced from our stochastic calculations, the values of
$\langle \cos \Phi \rangle_{\beta}$ could be extracted from those
thermal averages. Thus, we have that
\begin{equation}
\langle \cos \Phi \rangle_{\beta} =  \frac{(\delta/\Delta)
\mathrm{tanh} (\beta \Delta)}{\sqrt{1- (\epsilon/\Delta)^2
\mathrm{tanh}^2 (\beta \Delta)}}  .
\end{equation}
In Fig. \ref{fig2} we plot both behaviors as a function of the
temperature for $\delta = 1$ and two values of $\epsilon=0.5, 1$.
As can be seen, the agreement is fairly good. At very high
temperatures, where the classical limit is reached, the tail of
the coherence factor is given by
\begin{equation}
\langle \cos \Phi \rangle_{\beta} \simeq \pi^2
\frac{\delta}{\epsilon} I_1 (\beta \epsilon) \rightarrow
\frac{1}{T} .
\end{equation}
if the linear term is retained for the modified Bessel function of
first order. This expression is obtained approximately from Eq.
(\ref{F}) when $F(z,\Phi) = cos \Phi$.

%


Once the stochastic dynamics is well characterized, the next step
is to calculate the heat capacity. In thermodynamics, from the
knowledge of the partition function, the equilibrium
thermodynamical functions are also easily deduced such as the free
energy, the entropy, the heat capacity, etc. In particular, the
heat capacity at constant volume is expressed for a chiral system
as \cite{Bargueno2009}
\begin{equation}
C_v = k_B \beta^2 \Delta^2 \mathrm{sech}^2 \beta \Delta \label{Cv}
\end{equation}
displaying the so-called Schottky anomaly occurring in systems
with a limited number of energy levels. This thermodynamic
expression for the heat capacity is usually derived from one of
the two following expressions
\begin{eqnarray}
C_{v} & = &\frac{\partial U}{\partial T}
\nonumber \\
& = & k_B \beta^2 \frac{\partial ^2 Z}{\partial \beta^2}
\end{eqnarray}
%
where $U$ is the internal energy. For open systems, the coupling
to the heat bath defining the temperature is in general finite and
weak. The definition of the internal energy is not obvious.
Usually, one is inclined to assume that
\begin{equation}
U = \frac{\partial \langle E \rangle_{\beta}}{\partial T}
\end{equation}
where this energy is seen as the average energy of the chiral
system in the presence of the thermal bath and its mutual
interaction. In our context, it is the non--conserved energy given
by Eqs. (\ref{Hgamma}) and (\ref{Eave-t}). If we want to follow
the second expression, the partition function of the reduced
system has to be used. As pointed out previously
\cite{Hanggi2008,Ingold2009,Ingold2012}, these two routes may
differ yielding different results. In particular, the second route
leads to negative values at very low temperatures when dealing
with free damped particles. Specific heat anomalies in open
quantum systems are nowadays an important topic.

\begin{figure}[h!]
\includegraphics[width=0.45\textwidth,height=0.45\textwidth,angle=-90]
{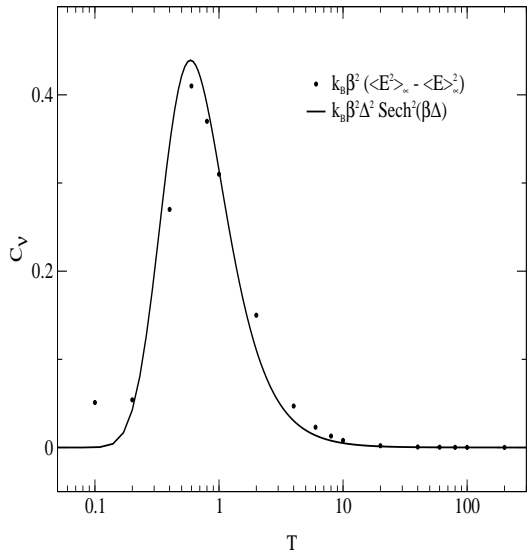} \caption{\label{fig4} Heat capacity of the chiral TLS
issued from the stochastic dynamics (solid points) and from the
thermodynamical equilibrium (solid line). Here the choice of
parameters are: $\delta = 1$, $\epsilon = 0.5$ and $\gamma = 0.1$
for $T > 1$ and 0.01 for $T < 1$ }
\end{figure}

\begin{figure}[h!]
\includegraphics[width=0.45\textwidth,height=0.45\textwidth,angle=-90]
{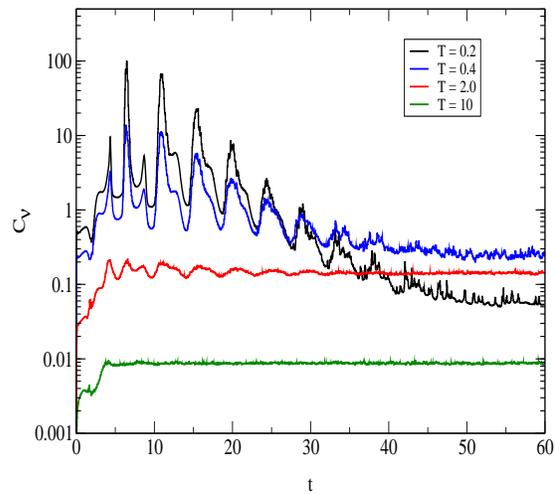} \caption{\label{fig5}  Time dependent energy
fluctuations expressed in terms of the heat capacity at four
different temperatures, covering a range around the critical value
which is $T_c= 1$ for $\delta = 1$ and $\epsilon =0.5$ $\gamma =
0.1$ for $T > 1$ and 0.01 for $T < 1$. }
\end{figure}

In this work, we are going to use a different strategy. The heat
capacity is evaluated from the energy fluctuations of the chiral
system as
\begin{equation}
C_{v} = \frac{1}{k_{B}T^{2}}\langle (E-\langle
E\rangle_{\beta})^{2}\rangle_{\beta} \label{Cv1}
\end{equation}
Following this procedure, the heat capacity is time--dependent
till it reaches a constant value at asymptotic times. In Fig.
\ref{fig4}, it is shown the heat capacity from Eq. (\ref{Cv})
(thermodynamical value, solid line) and from the energy
fluctuations through Eq. (\ref{Cv1}) after propagating the
equations of motion at asymptotic times (solid points). In Fig.
\ref{fig5}, the time evolution of the corresponding energy
fluctuations are also plotted for some different temperatures
displaying quite regular oscillations due to the interaction with
the bath and the tunneling process. As can be seen, the agreement
is fairly good taking into account the semi--log plot of the
horizontal axes.

The temperature behavior exhibited by the heat capacity displaying
the Schottky anomaly in Fig. \ref{fig4} leads to the definition of
a critical temperature defined by its maximum to be
\cite{Bargueno2009}
\begin{equation}
T_c \sim \frac{\Delta}{k_B 1.2}   .
\end{equation}
When $T > T_c$, the effect of $\epsilon$ is masked by thermal
effects which tend to wash out the population difference $z$
(racemization). On the contrary, for $T < T_c$, the value of the
ratio $\epsilon / \delta$ is critical. If this ratio is close to
unity, the whole dynamics is determined by the competition between
tunneling and asymmetry or bias. When it is much greater than one,
the tunneling process plays a minor role and the dynamics gives
place to localization. However, the racemization is always present
for ratios much less than one. Given the values of these
parameters in our calculation, the critical temperature is $T_c
\sim 1$ in units of $\Delta$. Thus, the range of temperatures
covered in all the calculations takes into account the coherent
and incoherent tunneling regimes.

Finally, several magnitudes can also be straightforward derived
from the energy fluctuations such as the geometric phase and the
diffraction pattern. In Ref. \cite{Dorta2012}, it was reported a
natural extension of the geometric phase to dissipative two level
systems. A further extension it is also possible for these systems
but the dynamics is stochastic. Thus, we have that
\begin{equation}
\phi_g \equiv \pi + \int_0^{\tau} H_{\gamma,\xi} (z,\Phi) dt
\end{equation}
where $\tau$ accounts for the period of a complete cycle. Due to
the fact that the quantity $H_{\gamma,\xi} (z,\Phi)$ is to be
erratic along time, it is better to average over the number of
total trajectories in order to have a smooth function with time,
the corresponding time integration being easily carried out
\begin{equation}
\langle \phi_g \rangle \equiv \pi + \int_0^{\tau} \langle
H_{\gamma,\xi} (z,\Phi) \rangle dt
\end{equation}

In a similar vein, information on interference experiments can be
straightforwardly extracted from the probability density of the
non-isolated TLS, that is,
\begin{equation}
I \propto | \Phi (t) | 1 + 2 |a_L (t)| |a_R (t)| cos \Phi (t)
\end{equation}
Now by using the effective Hamiltonian approach here developed,
the total intensity of the interference pattern for the Ohmic case
is given by
\begin{equation}
I \propto 1 + \frac{\epsilon}{z} - H_{\gamma,\xi} (z,\Phi)
\end{equation}

\section{Final discussion}

In this work, we have put on evidence that the stochastic dynamics
of a non-isolated chiral TLS, and issued from a classical
formalism, is able to reproduce the quantum termodynamical
functions such as partition function and heat capacity as well as
the so-called coherence factor. The corresponding classical
thermodynamical magnitudes calculated from the classical
Hamiltonian function are only valid at high temperatures. In other
words, we have here a case where a classical stochastic dynamics
is able to reproduce the coherent and incoherent regimes taken
place in a chiral TLS in presence of a thermal bath. After our
analysis, we are describing the coherence regime by means of a
classical dynamics. We think this is also a good example where the
origin of coherence effects (quantum or classical) here is not so
obvious, if not ambiguous \cite{Miller2012}. Finally, we have also
discussed the difference between the $\Phi$- and $z$-coupling in
this dynamics. As should be expected, the asymptotic behavior has
to be the same but the time evolution follows different paths.

\section{Appendix}

In this Appendix we derive the stochastic Langevin equations for a
two--level system when the system--bath coupling depends of the
system population difference instead of the system relative phase.
We will explicitly show that, for the dissipative case (that is,
for zero temperature), the magnitudes at the equilibrium are
independent of the type of coupling considered.

We start with the total Hamiltonian $H=H_{0}+H_{b}+H_{sb}$, where
$H_{0}$, given by Eq. (\ref{H}),
is the canonical Hamiltonian for the isolated two--level system,
\begin{equation}
H_{b}=\nobreak \frac{1}{2}\sum_{i} \left(\frac{p_{i}^{2}}{m_{i}}+m_{i}\omega_{i}^{2}x_{i}^{2} \right)
\end{equation}
is the Hamiltonian for the bath, which acts as a reservoir, and
can be represented as a set of harmonic oscillators, and
\begin{equation}
\label{coupling}
H_{sb}=\nobreak \sum_{i}\left[\frac{z^{2}d_{i}^{2}}{m_{i}\omega_{i}^{2}}-2 d_{i}z x_{i} \right]
\end{equation}
expresses an interaction term between the isolated system and the
bath, where $d_{i}$ are appropriate coupling constants. We note
that when the system--bath coupling is linear in $\Phi$, that is,
when the coupling is of the form $\Phi x_{i}$, the dynamics
corresponds to that employed in this article and in previous works
\cite{Bargueno2011,Dorta2012,chirality}. However, in this appendix
we investigate the effects of a $z x_{i}$ coupling, as shown in
Eq. (\ref{coupling}).

After the elimination of the bath degrees of freedom we arrive at
the following equations:
\begin{eqnarray}
\label{eqanomalous}
\dot z &=& -\sqrt{1-z^{2}}\sin \Phi \nonumber \\
\dot \Phi &=& \frac{z}{\sqrt{1-z^{2}}}\cos \Phi + \frac{\epsilon}{\delta}-
\int_{0}^{t}\gamma(t-s) \dot z ds + \xi(t),
\end{eqnarray}
where both the friction kernel and the noise function have the
usual definition \cite{chirality}. Therefore, the only difference
with the $\Phi x_{i}$ coupling lies at the dissipative $\gamma
\dot z$ term.

\begin{figure}[h!]
\includegraphics[width=0.5\textwidth,height=0.3\textwidth,angle=-90]
{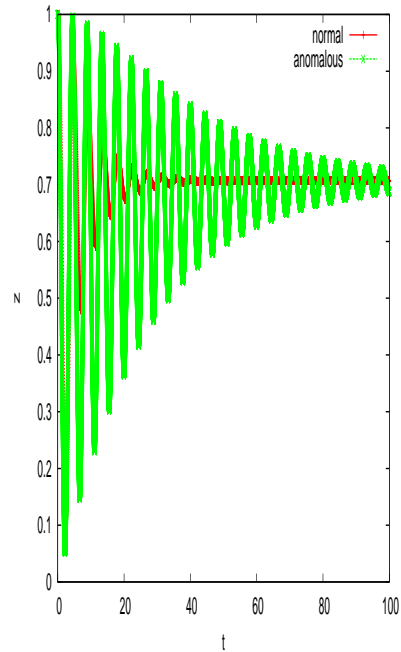} \caption{\label{fig1app} Comparison between
$\Phi$-coupling (red line) and $z$-coupling (green line) for the
population differences when $\epsilon=\delta=1$ and
$\gamma=10^{-1}$. See text for details.}
\end{figure}

To compare Eqs. (\ref{eqanomalous}) with their counterparts given
by Eq. (\ref{stochohm1}), we consider pure Ohmic dissipation at
zero temperature. In this case, the coupled equations are
\begin{eqnarray}
\label{eqanomalous}
\dot z &=& -\sqrt{1-z^{2}}\sin \Phi \nonumber \\
\dot \Phi &=& \frac{z}{\sqrt{1-z^{2}}}\cos \Phi + \frac{\epsilon}{\delta}-
\gamma\dot z.
\end{eqnarray}
\begin{figure}[t]
\includegraphics[width=0.3\textwidth,height=0.3\textwidth,angle=-90]
{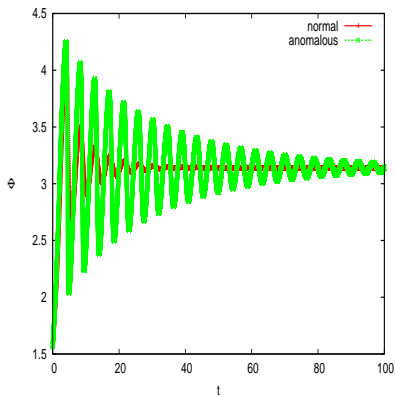} \caption{\label{fig2app} Comparison between
$\Phi$-coupling (red line) and $z$-coupling (green line)  for the
phase difference when $\epsilon=\delta=1$ and $\gamma=10^{-1}$.
See text for details.}
\end{figure}

\begin{figure}[t]
\includegraphics[width=0.3\textwidth,height=0.3\textwidth,angle=-90]
{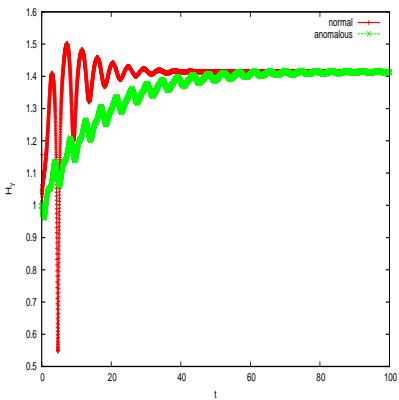} \caption{\label{fig3app} Comparison between
$\Phi$-coupling (red line) and $z$-coupling (green line) for the
energy when $\epsilon=\delta=1$ and $\gamma=10^{-1}$. See text for
details.}
\end{figure}

In Figs. (\ref{fig1app}),(\ref{fig2app}) and (\ref{fig3app}) we
plot the comparison between the population, phase differences, and
the energy fluctuations calculated with the normal ($\gamma \dot
\Phi$) and anomalous ($\gamma \dot z$) coupling terms for
$\epsilon = 1$ and $\gamma = 10^{-1}$. Apart from the relaxation
time and some differences in the amplitude of the oscillations,
both types of coupling give place to the same equilibrium values,
as expected.

\newpage

\noindent {\bf Acknowledgement} This work has been funded by the
MICINN (Spain) through Grant Nos. CTQ2008-02578, FIS2010-18132,
and by the Comunidad Aut\'onoma de Madrid, Grant No.
S-2009/MAT/1467. P. B. acknowledges a Juan de la Cierva fellowship
from the MICINN and A.D.-U. acknowledges a JAE fellowship from
CSIC. H. C. P.-R. and G. R.-L. acknowledge a scientific project
from INSTEC.

\end{document}